\documentstyle[twocolumn,prb,aps]{revtex}
\begin{document}
\title{Quantum Conductance of the  Single Electron Transistor}
\author{Xiaohui Wang}
\address{Fakult{\"a}t f{\"u}r Physik, Universit{\"a}t Freiburg, 
Hermann-Herder-Stra{\ss}e  3, 79104  Freiburg,  Germany}
\date{PRB 55, 12868 (1997)}
\maketitle
\begin{abstract}
The quantum conductance of the single-electron tunneling (SET) transistor 
is investigated in this paper by the functional integral approach.  The 
formalism is valid for arbitrary tunnel resistance 
of the junctions forming the SET transistor at any temperature.   The
path integrals are evaluated by the semiclassical method to yield 
an explicit non-perturbation form of the quantum conductance of the
SET transistor.  An anomaly of the quantum conductance is found if the
tunnel resistances are much smaller than the quantum resistance.  The
dependence of the conductance on the gate voltage is also discussed.

\end{abstract}


\narrowtext

Coulomb blockade effects in systems containing ultrasmall tunnel
junctions have attracted great interest in recent years
\cite{alt91,sz90,gra92}.
The basic device to show the transport properties of such structures
is the single electron tunneling transistor (SET-transistor), where an 
ultrasmall metallic island is formed by
two in-series connected tunnel junctions.  The island is coupled to a
gate voltage $V_G$ via a capacitance $C_G$.  A transport voltage $V$
is attached, say,  to the left lead electrode.  
The behavior of the SET-transistor has been investigated extensively
both theoretically and experimentally \cite{laf91,ave91,ful87,gee90}.
Most of the works have concentrated on the case of high tunnel
resistances and low temperatures in order to satisfy the conditions 
$R_T \gg R_K=h/e^2$ and $k_B T  \ll e^2/2C$ with $C$ of the same order of 
$C_G$  or capacitances of the tunnel junctions so that Coulomb blockade 
effects are obviously not washed out \cite{gra92}.  Under these conditions, 
however, the speed of controlable tunneling processes is strongly limited 
and  present devices must work at very low temperatures.  On the other hand, 
the latest experiments show that the  manifestation of Coulomb blockade at 
high temperatures is clearly observable, even for strong tunneling, and such 
effects are of significant applications \cite{pek94,est}.  Therefore it is  
meaningful to investigate such systems with various tunneling strengths 
and temperatures. 

The SET-transistor in the presence of an infinitesimal transport
voltage may be described by the Hamiltonian
$H=H_0+H_T$ \cite{sz90,sch94} ,
where 
$H_0=H_L+H_R+H_I+H_Q$ 
represents systems in the absence of tunneling.  Here $H_L$, $H_R$
and $H_I$ describe the free quasi-particles on the left lead
electrode, right lead electrode and on the island formed by the 
tunnel junctions, respectively.  $H_Q$ is the Coulomb energy of the
SET-transistor with the form 
$H_Q=E_c(q/e-n_{\rm ex})^2$, where  
$E_c=e^2/2C_\Sigma$
with the total capacitance 
$C_\Sigma=C_L+C_R+C_G$. 
The influence of the continuous change of the gate voltage is
described by $n_{\rm ex}=C_G V_G/e$ 
and $q$ is the charge operator on the island.

The tunneling Hamiltonian contains two parts describing tunneling
processes between the left lead electrode and the island 
\begin{eqnarray}
H_{T1}(\varphi_1) &=& H_{T1}^+(\varphi_1)+H_{T1}^-(\varphi_1) \nonumber \\ 
&=& \sum_{kq\sigma} (t_{kq\sigma}c^+_{k\sigma} c_{q\sigma} e^{-i\varphi_1} +
h.c.) 
\end{eqnarray}
and tunneling processes between the island and the right lead
electrode
\begin{eqnarray}
H_{T2}(\varphi_2) &=& H_{T2}^+(\varphi_2)+H_{T2}^-(\varphi_2) \nonumber \\
&=& \sum_{k'q\sigma} (t_{k'q\sigma}c^+_{k' \sigma} c_{q\sigma}
e^{-i\varphi_2} +h.c.). 
\end{eqnarray}
Here $k$, $k'$ and $q$ are the longitudinal wave numbers, $\sigma$
denotes the transversal and spin numbers, $\varphi_i$ is the phase
of the i-th tunnel junction conjugate to the charge of it $Q_i$.  The
conservation of
$\sigma$ during tunneling processes is included in the above
equations.  Since the typical impedance of the external
electromagnetic environment is much smaller than $R_K$, its influence on
the quantum conductance of the SET-transistor may normally be neglected 
\cite{dev90,wangl}.

Let us now introduce the total phase
$\Psi=\varphi_1+\varphi_2 =e \int_{-\infty}^{t} dt' V(t')$ 
conjugate to the total charge 
$Q=\kappa_1 Q_1+\kappa_2 Q_2$, 
and the phase 
$\varphi=\kappa_2 \varphi_1-\kappa_1 \varphi_2$ 
as a conjugate variable to the island charge $q$.  Here 
$\kappa_1=(C_R+C_G/2)/C_\Sigma$, and 
$\kappa_2=(C_L+C_G/2)/C_\Sigma$. 
Then the perturbation Hamiltonian as a linear function of the
transport voltage reads
\begin{equation} 
\delta H_T=\Psi F(\varphi)+{\cal O} (\Psi^2),
\end{equation} 
with the generalized force
\begin{equation} 
F(\varphi)=[\kappa_1 I_1(\varphi)-\kappa_2 I_2(\varphi)]/e,
\end{equation} 
where the tunnel current from the left lead electrode to the island is
\begin{equation} 
I_1(\varphi)=ie[H_{T1}^+(\varphi)-H_{T1}^-(\varphi)],
\end{equation}
and the one from the right lead electrode to the island reads 
\begin{equation} 
I_2(\varphi)=ie[-H_{T2}^+(-\varphi)+H_{T2}^-(-\varphi)] .
\end{equation} 

By employing Kubo's formula for the SET-transistor, we find that for 
$\tau >0$ its dc-conductance takes the following form
\begin{equation}
G_{dc}=\lim_{\omega \rightarrow 0} \omega^{-1} {\Im}m \{
\lim_{i\omega_l \rightarrow \omega+i\delta} e
\int_{0}^{\beta} d\tau e^{i\omega_l \tau} \langle I(\tau)F(0) \rangle \} ,
\end{equation}
where $\{ \omega_l \}$ are Matsubara frequencies with the definition 
$\omega_l=2\pi l/\beta, \ {\rm for} \  l=\pm 1, \pm2, \cdots$ and $I(\tau)$ 
may be chosen as $I_1(\tau)$ or $-I_2(\tau)$, because $-I_2$ 
flows in the direction that the transport voltage decreases. 
It will be shown that they lead to the same result.  

The undisturbed system is actually a single electron box (SEB) with
two in-parallel connected tunnel junctions\cite{laf91,sch94}.  The partition
function of this system may be described as a functional integral in the
following form \cite{sz90,ben83,wang96}
\begin{equation} 
Z=\int D \varphi e^{-S_\Sigma [\varphi]} ,
\end{equation} 
with the action
\begin{eqnarray}
 & & S_\Sigma [\varphi]=in_{\rm ex}[\varphi(\beta)-\varphi(0)]
+\int_0^\beta d \tau
\frac{\dot{\varphi}^2(\tau)}{4E_c} \nonumber \\
 & & -\int_0^\beta d \tau \int_0^\beta d \tau'
\alpha_\Sigma(\tau-\tau') \cos [\varphi(\tau)-\varphi(\tau')],  \label{act}
\end{eqnarray}
where 
$\alpha_\Sigma(\tau)=\alpha_1(\tau)+\alpha_2(\tau)$. 
Here $\alpha_i(\tau)$ is the damping kernel of the i-th tunnel
junction, which is an even function with a period
$\beta$ and its Fourier transform reads \cite{sz90,ben83,wang96}
\begin{equation}
\alpha_i(\tau)=\frac{1}{\beta} \sum_{l=-\infty}^{\infty} \alpha_i(\omega_l)
e^{i\omega_l \tau}, 
\end{equation}
where 
\begin{equation}
\alpha_i(\omega_l)=-\frac{\alpha_{ti}}{4\pi} |\omega_l|, \ \ \  \hbox{\rm for} 
\ \ \ |\omega_l| \ll D .
\end{equation}
Here 
$\alpha_{ti}=4\pi^2 t_i^2 \rho_i \rho'_i N_i$, 
which is related to the tunnel resistance through
$\alpha_{ti}=R_K/R_i$,
and thus 
$\alpha_{\Sigma}(\tau)=
\alpha_{t\Sigma} \alpha_i(\tau)/\alpha_{ti}$ 
with 
$\alpha_{t\Sigma}=\alpha_{t1}+\alpha_{t2}$ . 
The bandwidth in metals $D$ is normally much larger than
the single electron charging energy  $E_c$. 

The current-current correlation functions may also be described in a
path integral form by the generating functional approach as predicted by 
Ben-Jacob et al \cite{ben83,bro86}.  For the SET-transistor, we obtain for 
$i=1$ or $2$ 
\begin{eqnarray}
& & \langle I_i(\tau) I_i(0) \rangle=Z^{-1} \int D \varphi
e^{-S_\Sigma[\varphi]} \{ 
2e^2 \alpha_i(\tau) \nonumber \\
& & \times \cos [\varphi(\tau)-\varphi(0)] 
+I_{Ti}[\varphi, \tau] I_{Ti}[\varphi, 0] \} \label{auto}
\end{eqnarray}
and for $i\neq j$  
\begin{equation}
\langle I_i(\tau) I_j(0) \rangle=Z^{-1}  \int D \varphi e^{-S_\Sigma[\varphi]} 
I_{Ti}[\varphi, \tau] I_{Tj}[\varphi, 0],
\end{equation}
where 
\begin{equation}
I_{Ti}[\varphi, \tau]=-2e \int_{0}^{\beta} ds \alpha_i(\tau-s) \sin
[\varphi(\tau)-\varphi(s)].
\end{equation}

From (\ref{act}), we see that the action is a periodic function of
$\varphi$ with period $2\pi$, and thus the partition function
may be written as a sum over winding numbers, i.e.\ 
\begin{equation}
Z=\sum_{k=-\infty}^{\infty} \int_{\varphi(\beta)=\varphi(0)+2\pi k}  
D [\varphi] e^{-S_\Sigma[\varphi]},
\end{equation}
and so is the current auto-correlation function,
\begin{eqnarray}
& & \langle I_i(\tau) I_i(0) \rangle=Z^{-1}\sum_{k=-\infty}^{\infty}
\int_{\varphi(\beta)=\varphi(0)+2\pi k}  
D [\varphi] e^{-S_\Sigma[\varphi]} \nonumber \\
& & \{ 2 e^2 \alpha_i(\tau) \cos [\varphi(\tau)-\varphi(0)] 
+I_{Ti}[\varphi, \tau] I_{Ti}[\varphi, 0] \}.
\end{eqnarray}

Due to the nonlinear, nonlocal interactions described by the damping part 
of the action, the partition function and correlation functions cannot be 
evaluated exactly. For a single tunnel junction or the SEB, there are some 
methods such as renormalization group theory, sluggish phase transition 
technique, quasiclassical Langevin equation, self-consistent harmonic 
approximation and Monte Carlo simulation 
developed to evaluate such functional integrals approximately or numerically 
\cite{sz90,sch94,wang96,bro86,wang97,gol92,kos76,weg}. 
At not too low temperatures the  functional integrals 
may  well be evaluated systematically as a series of $\beta E_c$ by the 
semiclassical method including fluctuations beyond the Gaussian approximation 
as shown below.

From $\delta S_\Sigma[\varphi]=0$ and the boundary condition 
$\varphi(\beta)=\varphi(0)+2\pi k$ 
we get the trivial classical paths satisfying 
$\varphi_{\rm cl}^{(k)}(\tau)=\varphi(0)+\omega_k \tau$ 
with the classical action 
$S_{\rm cl}^{(k)}=\pi^2 k^2/\beta E_c+
|k|\alpha_{t\Sigma}/2$ .
For an arbitrary path 
$\varphi(\tau)=
\varphi_{\rm cl}^{(k)}(\tau)+\theta(\tau)$ 
of winding number $k$ fluctuating about the classical solution,
the second order variational action reads
\begin{eqnarray}
\delta^2 S^{(k)} &=& \int_{0}^{\beta} d\tau 
\frac{\dot{\theta}^2}{4E_c}+\frac12
\int_{0}^{\beta} d\tau \int_{0}^{\beta} 
d\tau'\alpha_\Sigma(\tau-\tau') \nonumber \\
& & \times \cos[\omega_k(\tau
-\tau')][\theta(\tau)-\theta(\tau')]^2. 
\end{eqnarray}
In terms of the Fourier transform 
$\theta(\tau)=\sum_{l=-\infty}^{\infty} \theta_l e^{i \omega_l \tau}$, 
with $\theta_l=\theta'_l+i\theta''_l$ and  
$\theta_{-l}=\theta_l^*$, one has   
\begin{equation}
\delta^2S^{(k)}= \sum_{l=1}^{\infty} 
\lambda_l^{(k)} ({\theta'_l}^2+{\theta''_l}^2) , 
\end{equation}
where for $l \ll \beta D$, the eigenvalues read 
\begin{equation}
\lambda_l^{(k)}= \frac{\omega_l^2}{2 E_c}+\Theta(l-|k|) 
\frac{\alpha_{t\Sigma} \, \omega_{l-|k|}}{2 \pi}  .
\end{equation}
At high temperatures the eigenvalues are large and the trivial 
classical paths are stable.   
For very large $l$ the eigenvalues 
approach  $\omega_l^2/2 E_c$
 independent of $k$.  Upon  normalization of the  path
integral to the one of $k=0$, the large $l$ contributions cancel. It is then 
straightforward to show that
\begin{equation}
Z_{\rm semi}=Z_0 \left[ 1+2 \sum_{k=1}^{\infty} C_k \cos (2\pi k n_{\rm ex})  
\right] , 
\end{equation}
where  
$Z_0=\prod_{l=1}^{\infty} \pi/\beta \lambda_l^{(0)}$ 
is the contribution of paths with winding number 0 and
\[
C_k=\frac{\Gamma (1+k_+) \Gamma (1+k_-)}{\Gamma^2 (1+k) 
\Gamma (1+\mu_\Sigma)} e^{-S_{cl}^{(k)}} , 
\]
where 
$\mu_\Sigma=\alpha_{t\Sigma}\beta E_c/2\pi^2$ 
and 
$k_{\pm}=k+\mu_\Sigma/2 \pm \sqrt{4\mu_\Sigma  k+\mu_\Sigma^2}/2$.  

The current-current correlation functions may be evaluated in the same way.  
In order to show the evaluation explicitly, we calculate here only $k=0$
terms in detail, which are the dominating terms if the  
temperatures are not too low. Since for $\beta E_c \leq 1$, the semiclassical
approach is well-behaved, the $k \neq 0$ terms are depressed at least by a
factor of $\exp (-\pi^2/\beta E_c)$, and thus much smaller than the $k=0$
terms.   In this case the current auto-correlation function is
approximately of the following form 
\begin{eqnarray}
& & \langle I_i(\tau) I_i(\tau') \rangle_f^{(0)}=2e^2 \alpha_i(\tau-\tau') 
Z_{\rm semi}^{-1}
\int D \theta e^{-\delta^2 S[\theta]} \nonumber \\
& & \{ 1-\frac{1}{2} [\theta(\tau)-\theta(\tau')]^2 \}. 
\end{eqnarray}
In the Fourier space, we have 
\begin{eqnarray}
& &\int D \theta e^{-\delta^2 S[\theta]} 
\{ 1-\frac{1}{2} [\theta(\tau)-\theta(\tau')]^2 \} \nonumber \\
& &=Z_0 \left\{
1-2\sum_{l=1}^{\infty}\frac{1}{\beta \lambda_l^{(0)}}+2\sum_{l=1}^{\infty}
\frac{\cos \omega_l (\tau-\tau')}{\beta \lambda_l^{(0)}} \right \}.
\end{eqnarray}
Hence the corresponding conductance may be evaluated to give
\begin{eqnarray}
& & G_{i,f}^{(0)}=\lim_{\omega \rightarrow 0} \omega^{-1} {\Im}m \{
\lim_{i\omega_l \rightarrow \omega+i\delta} 
\int_{0}^{\beta} d\tau e^{i\omega_l \tau} \langle I_i(\tau)I_i(0)
\rangle_f^{(0)} \} \nonumber \\ 
& & =\lim_{\omega \rightarrow 0} \omega^{-1}{\Im}m 
\left\{
\lim_{i\omega_l \rightarrow \omega+i\delta} (2e^2)  
\left[ \alpha_i(\omega_l) -2 \sum_{n=1}^{\infty}
\frac{\alpha_i(\omega_l)}{\beta \lambda_n^{(0)}} 
\right. \right. \nonumber \\
& & \left. \left. 
+\sum_{m=1}^{\infty}
\frac{\alpha_i(\omega_m+\omega_l)+\alpha_i(\omega_m-\omega_l)}{\beta
\lambda_m^{(0)}} \right] \right \}.
\label{g1} 
\end{eqnarray}
The summations in the above equation may be carried out and finally 
we obtain 
\begin{equation}
G_{i,f}^{(0)}=D_{01}/R_i
\end{equation}
where the prefactor including the leading correction from fluctuations reads
\begin{equation} 
D_{01}=1-2 [\Psi(\mu_\Sigma+1)-\Psi(1)]/\alpha_{t\Sigma}-\beta E_c
\Psi'(\mu_\Sigma+1)/\pi^2. \label{d01}
\end{equation}
Here the definition of the quantum resistance $R_K$
has been used to get the final result.  Note that  we have taken $\hbar=1$
in this paper for convenience.  

The second term of the current auto-correlation function and the
correlation function between currents through different tunnel
junctions may be evaluated in the same manner.  After some algebra we find 
that the terms from the correlation function between the currents of the 
same junction satisfy 
\begin{eqnarray}
G_{ii}^{(0)}&=&\lim_{\omega \rightarrow 0} \omega^{-1} {\Im}m \{
\lim_{i\omega_l \rightarrow \omega+i\delta} \langle I_i(\tau) I_i(0)
\rangle^{(0)} \} \nonumber \\
&=& \frac{D_{01}}{R_i} \left[1-\frac{R_1 R_2}{R_i (R_1+R_2)} \right] ,  
\end{eqnarray}
and the one between different tunnel junctions is of the following form 
\begin{eqnarray}
G_{ij}^{(0)} &=& \lim_{\omega \rightarrow 0} \omega^{-1} {\Im}m \{
\lim_{i\omega_l \rightarrow \omega+i\delta} \langle I_i(\tau) I_j(0)
\rangle^{(0)}  \} \nonumber \\
&=& -\frac{D_{01}}{R_1+R_2} .
\end{eqnarray}
The total dc-conductance of the SET-transistor may be evaluated accordingly.
The quantum conductance observed from the left tunnel junction reads
\begin{eqnarray}
G_{dc}^{(L)}&=&\lim_{\omega \rightarrow 0} \omega^{-1} {\Im}m \{
\lim_{i\omega_l \rightarrow \omega+i\delta} e
\int_{0}^{\beta} d\tau e^{i\omega_l \tau} \langle I_1(\tau)F(0)
\rangle^{(0)} \}
\nonumber \\
&=& \kappa_1 G_{11}^{(0)}-\kappa_2 G_{12}^{(0)} \nonumber \\
&=& \frac{D_{01}}{R_1+R_2} . \label{gl}
\end{eqnarray}
Likewise, the conductance observed from the right junction takes the
following form
\begin{eqnarray}
G_{dc}^{(R)}&=&-\lim_{\omega \rightarrow 0} \omega^{-1} {\Im}m \{
\lim_{i\omega_l \rightarrow \omega+i\delta} e
\int_{0}^{\beta} d\tau e^{i\omega_l \tau}  \langle I_2(\tau)F(0)
\rangle^{(0)} \}
\nonumber \\
&=& -\kappa_1 G_{21}^{(0)}+\kappa_2 G_{22}^{(0)} \nonumber \\
&=& \frac{D_{01}}{R_1+R_2} , \label{gr}
\end{eqnarray}
which is exactly the same as $G_{dc}^{(L)}$.
  
For $\mu_\Sigma \ll 1$, the factor $D_{01}$ may be simplified to give
\begin{equation} 
D_{01}^{(n)}=1-\beta E_c/3+{\cal O} [(\beta E_c)^2] .
\end{equation}
In the high temperature limit, the ohmic behavior 
$G_{ohm}=1/(R_1+R_2)$ 
is recovered.  Up to the order of $\beta E_c $, this result is in good
agreement with the one by exploiting Fermi golden rule 
\cite{pek94,est,dev90}. Note that (\ref{gl}) or (\ref{gr}) are also
valid  for the SET-transistor formed by tunnel junctions with different 
resistances and capacitances, while by directly taking the tunnel Hamiltonian 
as a perturbation to calculate the quantum conductance analytically the 
junctions are taken to be identical \cite{pek94,est}.

An anomaly of the quantum conductance occurs if one or both of the tunnel 
resistances are much smaller than $R_K$ so that $\alpha_{t\Sigma} \gg
\pi^2/\beta E_c \gg 1$. Since the form of $D_{01}$ holds only if $\beta E_c$  
is very small, we have then 
\begin{equation} 
D_{01}^{(a)}=1-2 [ \ln (\alpha_{t\Sigma})+ \ln (\beta E_c)]/\alpha_{t\Sigma} 
+{\cal O}(1/\alpha_{t\Sigma}).
\end{equation}
This anomalous relation may be observed only under the above-mentioned 
strict conditions, the calculation here, however, shows that the linear 
dependence of the leading correction of the quantum conductance on 
$\beta E_c$ is generally not valid 
if $\alpha_{t\Sigma} > \pi^2/\beta E_c $.

The quantum conductance of the SET-transistor as a series of $\beta E_c$ may 
be evaluated further if higher orders of the
expansion of the action,  $\cos [\theta(\tau)-\theta(\tau')]$ and
$\sin [\theta(\tau)-\theta(\tau')]$  in (\ref{auto}) are taken into account.  
To the order of $(\beta E_c)^2$ we obtain that the
quantum  conductance of the SET-transistor for $\alpha_{t\Sigma} <
1/\beta E_c$ takes the same form as (\ref{gl}) or (\ref{gr}), yet with a 
prefactor 
\begin{equation} 
D_{02}=D_{01}+[0.0667+0.0185\alpha_{t \Sigma}] (\beta E_c)^2 
+{\cal O} [(\beta E_c)^3]  \label{d02}
\end{equation}
instead of $D_{01}$. 
Higher orders of $\beta E_c$ may also be determined in the same way.
It will  be very tedious to calculate such terms and a finite number
of terms  in the series is probably not enough to predict the
behavior of  the SET-transistor at low temperatures, because the
series is an  asymptotic one.  At low temperatures more sophisticated  
techniques are needed to evaluate the path integrals instead of the
simple semiclassical method 
\cite{sz90,sch94,wang96,bro86,wang97,gol92,kos76,weg}. 

The quantum conductance of the SET-transistor is an oscillating function of
the gate voltage if the terms with $k \neq 0$ are taken into account. 
Then the quantum conductance of the SET-transistor as a function of the
gate voltage is found to be 
\begin{equation}
G_{dc}=\frac{D_0+2C_1 D_1 \cos (2\pi n_{\rm ex})}{R_1+R_2},
\end{equation}
where at high temperatures the explicit form of $D_0$ is given by (\ref{d01}) 
or (\ref{d02}) and the leading term of $D_1$ reads 
\begin{equation}
D_{1,l}=-4\pi^2/\beta E_c-\alpha_{t\Sigma}+2+2\pi^2/3 
+{\cal O}(\beta E_c). 
\end{equation}

In summary, quantum conductance of the SET-transistor as a function of the gate
voltage has been discussed in this work.  By employing the functional
integral approach we have obtained the general formula of the quantum
conductance, which is valid for arbitrary tunnel resistance of the
junctions forming the SET-transistor at any temperature.  
At not too low temperatures the path integrals have
been evaluated by the semiclassical method to yield an explicit form of
the quantum conductance.  This result gives clear criterions to the
application of the SET-transistor as e.g.\ a new kind of thermometers
insensitive to magnetic field by taking advantage of the
linear dependence of the quantum conductance on $\beta E_c$
\cite{pek94}.  We have also proved that under typically experimental 
conditions the linear dependence of the quantum 
conductance on $\beta E_c$ is also valid for the SET-transistor formed by 
tunnel junctions with different parameters.  If the tunnel resistances 
are much smaller the quantum resistance, an anomaly of the quantum 
conductance has also been studied.  
Our treatment is nonperturbative and includes also the dependence of
the quantum conductance on the gate voltage.  At low temperatures 
the trivial classical paths 
are no more stable, thus the derived path integrals must be evaluated by 
methods other than the simple semiclassical approach so that  
large quantum fluctuations are properly treated.  

The author would like to thank M.\ B{\"u}ttiker, D.\ Esteve, G.\  
G{\"o}ppert and P.\ Joyez for fruitful 
discussions and the Deutsche Forschungsgemeinschaft for financial support.

\end{document}